\documentclass[a4paper,10pt,twoside]{cpc-hepnp}

\usepackage{multicol}
\usepackage{graphicx}
\usepackage{booktabs}
\usepackage{amssymb,bm,mathrsfs,bbm,amscd}
\usepackage[tbtags]{amsmath}
\usepackage{lastpage}

\begin{document}
%\begin{CJK*}{GBK}{song}

\fancyhead[co]{\footnotesize Zhong-Cheng Yang~ et al: The possible hidden-charm molecular baryons ...}

\footnotetext[0]{Received 14 March 2009}

\title{The possible hidden-charm molecular baryons composed of
anti-charmed meson and charmed baryon\thanks{This project is supported by the National Natural Science
Foundation of China under Grants No. 11175073, 11075004, 11021092,
11035006, 11047606 and, 10805048, the Ministry of Science
and Technology of China (No. 2009CB825200), and the Ministry of
Education of China (FANEDD under Grant No. 200924, DPFIHE under
Grants No. 20090211120029, NCET under Grant No. NCET-10-0442, the
Fundamental Research Funds for the Central Universities under
Grant No. lzujbky-2010-69).
}}

\author{%
      Zhong-Cheng Yang$^{1;}$%
      \quad
      Zhi-Feng Sun$^{2,4)}$
     \quad Jun He$^{1,3;1)}$\email{junhe@impcas.ac.cn}%
    \quad Xiang Liu$^{2,4;2)}$\email{xiangliu@lzu.edu.cn}
\quad Shi-Lin Zhu$^{1;3)}$\email{zhusl@pku.edu.cn}
}
\maketitle

\address{%
$^1$ Department of Physics and State Key Laboratory of Nuclear Physics and Technology,
Peking University, Beijing 100871, China\\
%Institution or University where the author works,  district,  postal code,  country\\
$^2$Research Center for Hadron and CSR Physics, Lanzhou University and Institute of Modern Physics of CAS, Lanzhou 730000, China\\
$^3$Nuclear theory group, Institute of Modern Physics of CAS, Lanzhou 730000, China\\
$^4$Schiool of Physical Science and Technology, Lanzhou
University, Lanzhou 730000, China
% {\bf Example}: Institute of High Energy Physics, Chinese Academy of Sciences, Beijing 100049, China\\
}

\begin{abstract}
With the one-boson-exchange model we have studied the possible
existence of the very loosely bound hidden-charm molecular baryons
composed of anti-charmed meson and charmed baryon. Our numerical
results indicate that there exist $\Sigma_c\bar{D}^*$ states with $I(J^P)=\frac{1}{2}(\frac{1}{2}^-), \frac{1}{2}(\frac{3}{2}^-), \frac{3}{2}(\frac{1}{2}^-), \frac{3}{2}(\frac{3}{2}^-)$ and $\Sigma_c\bar{D}$ state with $\frac{3}{2}(\frac{1}{2}^-)$. But the $\Lambda_c \bar{D}$ and
$\Lambda_c \bar{D}^*$ molecular states do not exist.
\end{abstract}


\begin{pacs}
14.20.Pt, 12.40.Yx, 12.39.Hg
%1---3 PACS(Physics and Astronomy Classification Scheme, http://www.aip.org/pacs/pacs.html/)
\end{pacs}

\begin{multicols}{2}

\section{Introduction}\label{sec1}

In the past eight years, more and more experimental observations
of new hadron states were announced, which has inspired extensive
interest in revealing the underlying structure of these newly
observed states. Besides making the effort to categorize them
under the framework of the conventional $q\bar{q}$ or $qqq$
states, theorists have also tried to explain some of these newly
observed hadrons as exotic states due to their peculiarities
different from the conventional $q\bar{q}$ or $qqq$ state.

\begin{center}
\tabcaption{The thresholds near the corresponding newly observed
hadron states. \label{threshold}}
\footnotesize
\begin{tabular*}{80mm}{c@{\extracolsep{\fill}}ccc}
\toprule Observation & Threshold &Observation & Threshold \\
\hline
$X(1860)$\cite{Bai:2003sw}&$p\bar{p}$&$D_s(2317)$\cite{Aubert:2003fg}&$DK$\\
$D_s(2460)$\cite{Besson:2003cp}&$D^*K$&$X(3872)$\cite{Choi:2003ue}&$D^*D$\\
$Y(3940)$\cite{Abe:2004zs}&$D^*D^*$&$Y(4140)$\cite{Aaltonen:2009tz}&$D_s^*D_s^*$\\
$Y(4274)$\cite{Aaltonen:2011at}&$D_s(2317)D$&$Y(4630)$\cite{Pakhlova:2008vn}&$\Lambda_c\Lambda_c$\\
$Z^+(4430)$&$D_1D^*/D_1^\prime D^*$&$Z^{+}(4250)$\cite{Mizuk:2008me}&$D_1D/D_0D^*$\\
$\Lambda_c(2940)$ \cite{Aubert:2006sp}&$D^*N$&$\Sigma_c(2800)$\cite{Mizuk:2004yu}&$DN$\\
\bottomrule
\end{tabular*}
\end{center}

Among different schemes to explain the structures of these newly
observed hadrons, molecular states composed of a hadron pair
become a very popular one due to the fact that the corresponding
observations are often near the threshold of a pair of hadrons as
in Table \ref{threshold}. In order to explore whether these newly
observed hadrons can be accommodated in the molecular framework,
there are many theoretical calculations of various molecular
states
\cite{Liu:2004er,He:2006is,Liu:2007ez,Liu:2007bf,Liu:2007fe,Tornqvist:2003na,Swanson:2004pp,Liu:2008fh,
Close:2009ag,Close:2010wq,Lee:2009hy,Xu:2010fc,
Liu:2008du,Liu:2008xz,
Liu:2008tn,Liu:2009ei,Hu:2010fg,Shen:2010ky,He:2010zq,Liu:2010hf,Liu:2008mi,Liu:2009zzf,
Liu:2009wb,Ding:2007ar,Ding:2008mp,Ding:2008gr,Ding:2009zq,Lee:2011rk,Chen:2011ct}.

Generally speaking, conventional hadrons with a charm quark can be
grouped into three families, i.e., charmonium, charmed meson,
charmed baryon with the configurations $[c\bar{c}]$, $[c\bar{q}]$,
$[cqq]$ respectively, where $q$ or $\bar{q}$ denotes the light
quark or anti-quark with different flavors. In principle, we may
extend these configurations by adding ${q\bar{q}}$ pair, which is
allowed by Quantum Chromodynamics (QCD). Such extension results in
three new exotic configurations $[c\bar{c}{q\bar{q}}]$,
$[c\bar{q}{q\bar{q}}]$, $[cqq{q\bar{q}}]$, which can be named as
molecular charmonium, moleular charmed meson and molecular charmed
baryon respectively if the corresponding constituents in these
configurations are color singlet. Inspired by the recent
experimental observations, many theoretical investigations
focusing on molecular charmonium, moleular charmed meson and
molecular charmed baryon have been performed\cite{He:2006is,Liu:2007ez,Liu:2007bf,Liu:2007fe,Tornqvist:2003na,Swanson:2004pp,Liu:2008fh,
Close:2009ag,Close:2010wq,Lee:2009hy,Xu:2010fc,Liu:2008du,Liu:2008xz,
Liu:2008tn,Liu:2009ei,Hu:2010fg,Shen:2010ky,He:2010zq,Liu:2010hf,Liu:2008mi,Liu:2009zzf,
Liu:2009wb,Ding:2007ar,Ding:2008mp,Ding:2008gr,Ding:2009zq,Lee:2011rk,Chen:2011ct}.

Apart from the above exotic molecular systems discussed
extensively in literatures, there may also exist new
configurations of the exotic molecular state if adding $qqq$ into
$[c\bar{c}]$ and $[cqq]$, which correspond to the exotic molecular
states with components $[c\bar{c}qqq]$ and $[cqqqqq]$. These
states may be accessible by future experiments such as PANDA,
Belle II and SuperB, etc, since the masses of the lightest exotic
molecular states with components $[cqqqqq]$ and $[c\bar{c}qqq]$
are just about 3.3 GeV and 4.1 GeV respectively.

At present, carrying out the dynamical study of these exotic
molecular systems becomes especially important, which will provide
experimentalists with valuable information such as their mass
spectrums and decay behaviors. There have been lots of theoretical
work recently. In Ref. \cite{Liu:2011xc}, Liu and Oka discussed
whether there exist the $\Lambda_cN$ molecular states. Narrow
$N^*$ and $\Lambda^*$ resonances with hidden charm were proposed
as the meson-baryon dynamically generated states\cite{Wu:2010jy}.
Later, the authors in Ref. \cite{Wang:2011rg} calculated the
S-wave $\Sigma_c \bar{D}$ and $\Lambda_c \bar{D}$ states with
isospin $I=1/2$ and spin $S=1/2$ using the chiral constituent
quark model and the resonating group method.

In this work, we will investigate the hidden-charm molecular
baryons which are composed of a S-wave anti-charmed meson and an
S-wave charmed baryon. The S-wave charmed baryons can be assigned
as either the symmetric $6_F$ or antisymmetric $\bar{3}_F$ flavor
representation as illustrated in Fig. \ref{mb}. Thus, the
spin-parity of the S-wave charmed baryons is $J^P=1/2^+$ or
$3/2^+$ for $6_F$ and $J^P=1/2^+$ for $\bar{3}_F$. The
pseudoscalar and vector anti-charmed mesons constitute an S-wave
anti-charmed mesons. In the following, we mainly focus on the
hidden-charm molecular states composed of these charmed baryons
and anti-charmed mesons existing in the green range.

\begin{center}
\includegraphics[width=8cm]{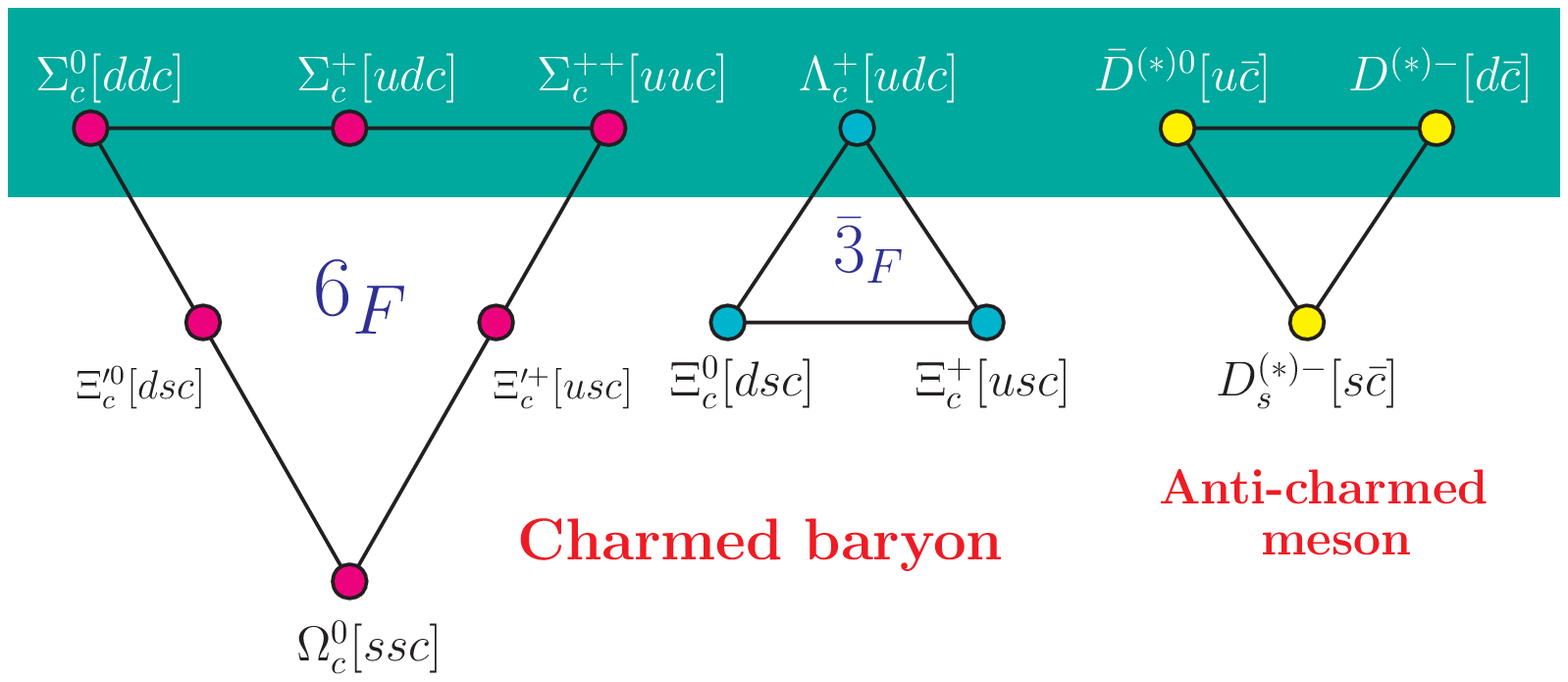}
\figcaption{(Color online).
The S-wave charmed baryons with $J^P=1/2^+$ and the S-wave
anti-charmed pseudoscalar/vector mesons contributing to the
double-charm molecular baryons. \label{mb}}
\end{center}

We apply the one boson exchange (OBE) model to study the
hidden-charm molecular states, which is an effective approach for
calculating the hadron-hadron interaction\cite{Tornqvist:2003na,Liu:2008du,Liu:2008xz,
Liu:2008tn,Liu:2009ei,Hu:2010fg,Shen:2010ky}. The interactions
between S-wave anti-charmed meson and S-wave charmed baryon with
$J^P=1/2^+$ are described in terms of the meson exchange with
phenomenologically determined parameters. In our former work\cite{He:2010zq}, we once studied the interaction between the
vector charmed meson $D^*$ and nucleon $N$, which could be related
to $\Lambda_c(2940)$\cite{Aubert:2006sp}. To some extent, the
framework in this work is similar to that in Ref.
\cite{He:2010zq}.

This paper is organized as follows. After the introduction, we
present the calculation of interactions between S-wave
anti-charmed meson and S-wave charmed baryon with $J^P=1/2^+$. In Sec. 3, the numerical results are presented. The last section is the discussion and conclusion.

\section{The interaction of hidden-charm molecular baryons}\label{sec2}

\subsection{Flavor wave functions}

In this work, we mainly focus on the systems composed of an S-wave
anti-charmed meson and an S-wave charmed baryon with $J^P=1/2^+$.
These systems are of negative parity. Furthermore, the states
composed of an S-wave charmed baryon with $J^P=1/2^+$ and an
anti-charmed meson with spin zero include $\bar{D}\Lambda_c$ and
$\bar{D}\Sigma_c$ systems, which are of
$I(J^P)=0(\frac{1}{2}^-),\,\frac{1}{2}(\frac{1}{2}^-),\,\frac{3}{2}(\frac{1}{2}^-)$.
Such a system contains one state only
\begin{eqnarray}
    \Big|I\Big(\frac{1}{2}^-\Big)\Big\rangle_0:&&\ \
    \Big|^2\mathbb{S}_{\frac{1}{2}}\Big\rangle.\label{Eq:wf01}
\end{eqnarray}
For comparison, $\bar{D}^*\Lambda_c$ and $\bar{D}^*\Sigma_c$ are
the systems with an S-wave charmed baryon with $J^P=1/2^+$ and an
anti-charmed meson with spin one, which are of
$I(J^P)=\frac{1}{2}(\frac{1}{2}^-),\,\frac{3}{2}(\frac{1}{2}^-),\,\frac{1}{2}(\frac{3}{2}^-),\,\frac{3}{2}(\frac{3}{2}^-)$.
Thus, several states may contribute to such systems
\begin{eqnarray}
    \Big|I\Big(\frac{1}{2}^-\Big)\Big\rangle_1:&&\ \     \Big|^2\mathbb{S}_{\frac{1}{2}}\Big\rangle,\ \ \ \
\Big|^4\mathbb{D}_{\frac{1}{2}}\Big\rangle.\label{Eq:wf1}\\
\Big|I\Big(\frac{3}{2}^-\Big)\Big\rangle_1:&&\ \
\Big|^4\mathbb{S}_{\frac{3}{2}}\Big\rangle,\ \ \ \
    \Big|^2\mathbb{D}_{\frac{3}{2}}\Big\rangle,\ \ \ \
    \Big|^4\mathbb{D}_{\frac{3}{2}}\Big\rangle.
    \label{Eq:wf3}
\end{eqnarray}
In Eqs. (\ref{Eq:wf01})-(\ref{Eq:wf3}), we use notation
$^{2S+1}L_J$ to distinguish different states, where $S$, $L$ and
$J$ denote the total spin, angular momentum and total angular
momentum respectively. Indices $\mathbb{S}$ and $\mathbb{D}$ show
that the couplings between anti-charmed meson and charmed baryon
occur via the $S$-wave, $D$-wave interactions respectively.

The general expressions of states in Eq.~(\ref{Eq:wf01}) and
Eqs.~(\ref{Eq:wf1})-(\ref{Eq:wf3}) can be explicitly written as
\begin{eqnarray}
    \Big|^{2S+1}L_J\Big\rangle_0    &=&\chi_{\frac{1}{2}M}Y_{00},\label{1}\\
    \Big|^{2S+1}L_J\Big\rangle_1    &=&\sum_{m,m',m_L,m_S}C_{Sm_S,Lm_L}^{JM}
    C_{\frac{1}{2}m,1m'}^{Sm_S}\nonumber\\&&\times
    \epsilon^{m'}_{n}\chi_{\frac{1}{2}
    m}Y_{Lm_L},\label{2}\label{po}
\end{eqnarray}
where $C_{\frac{1}{2}m,Lm_L}^{JM}$, $C_{Sm_S,Lm_L}^{JM}$ and
$C_{\frac{1}{2}m,1m'}^{Sm_S}$ are Clebsch-Gordan coefficients.
$Y_{Lm_L}$ is the spherical harmonics function. $\chi_{\frac{1}{2}
m}$ denotes the spin wave function. The polarization vector for
$\bar{D}^*$ is defined as
$\epsilon^m_\pm=\mp\frac{1}{\sqrt{2}}(\epsilon^m_x\pm
i\epsilon^m_y)$ and $\epsilon^m_0=\epsilon^m_z$.

\subsection{Effective Lagrangian}\label{sub1}

When adopting the OBE model to calculate the effective potential
of the hidden-charm molecular baryons, we need to construct the
effective Lagrangian describing the interactions of the charmed or
anti-charmed baryons/mesons with the light mesons
($\pi,\eta,\rho,\omega,\sigma,\cdots$). According to the chiral
symmetry and heavy quark limit, the Lagrangian for the S-wave
heavy mesosns interacting with light pseudoscalar, vector and
vector mesons reads\cite{Cheng:1992xi,Yan:1992gz,Wise:1992hn,Burdman:1992gh,Casalbuoni:1996pg}
\begin{eqnarray}
\mathcal{L}_{HH\mathbb{P}}&=& ig_{1}\langle \bar{H}_a^{\bar{Q}} \gamma_\mu
A_{ba}^\mu\gamma_5 {H}_b^{\bar{Q}}\rangle,\label{eq:lag}\\
\mathcal{L}_{HH\mathbb{V}}&=& -i\beta\langle \bar{H}_a^{\bar{Q}}  v_\mu
(\mathcal{V}^\mu_{ab}-\rho^\mu_{ab}){H}_b^{\bar{Q}}\rangle
\nonumber\\&&+i\lambda\langle \bar{H}_b^{\bar{Q}}
\sigma_{\mu\nu}F^{\mu\nu}(\rho)\bar{H}_a^{\bar{Q}}\rangle
,\\
\mathcal{L}_{ HH\sigma}&=&g_s \langle \bar{H}_a^{\bar{Q}}\sigma
\bar{H}_a^{\bar{Q}}\rangle,\label{eq:lag2}
\end{eqnarray}
which satisfies Lorentz and $C$, $P$, $T$ invariance, where
$\langle\cdots\rangle$ denotes the trace over the the $3\times 3$
matrices. The multiplet field $H$ composed of the pseudoscalar
$\mathcal{P}$ and vector $\mathcal{P}^*$ with
${\mathcal{P}}^{(*)T} =(D^{(*)0},D^{(*)+},D_s^{(*)+})$ or
$(B^{(*)-},\bar{B}^{(*)0},\bar{B}_s^{(*)0})$ is defined as
$H_a^{\bar{Q}}=[\tilde{\mathcal{P}}^{*\mu}_a\gamma_\mu-\tilde{\mathcal{P}}_a\gamma_5]\frac{1-\rlap\slash
v}{2}$ and
$\bar{H}=\gamma_0H^\dag\gamma_0$ with $v=(1,\mathbf{0})$. The
$\tilde{\mathcal{P}}
$ and $\tilde{\mathcal{P}}^*%(\widetilde{\mathcal{P}}^*)
$ satisfy the normalization relations $\langle
0|\tilde{{\mathcal{P}}}|\bar{Q}{q}(0^-)\rangle
%=\langle 0|\widetilde{\mathcal{P}}|\bar{Q}q(0^-)\rangle
=\sqrt{M_\mathcal{P}}$ and $\langle
0|\tilde{{\mathcal{P}}}^*_\mu|\bar{Q}{q}(1^-)\rangle=%\langle
%0|\widetilde{\mathcal{P}}^{*}_\mu|\bar{Q}q(1^-)\rangle
\epsilon_\mu\sqrt{M_{\mathcal{P}^*}}$. In the above expressions,
the axial current is
$A^\mu=\frac{1}{2}(\xi^\dag\partial_\mu\xi-\xi \partial_\mu
\xi^\dag)=\frac{i}{f_\pi}\partial_\mu{\mathbb P}+\cdots$ with
$\xi=\exp(i\mathbb{P}/f_\pi)$ and $f_\pi=132$ MeV.
$\rho^\mu_{ba}=ig_{V}\mathbb{V}^\mu_{ba}/\sqrt{2}$,
$F_{\mu\nu}(\rho)=\partial_\mu\rho_\nu - \partial_\nu\rho_\mu +
[\rho_\mu,{\ } \rho_\nu]$, and $g_V=m_\rho/f_\pi$. Here, $\mathbb
P$ and $\mathbb V$ are the pseudoscalar and vector matrices
\begin{eqnarray}
    {\mathbb P}&=&\left(\begin{array}{ccc}
        \frac{1}{\sqrt{2}}\pi^0+\frac{\eta}{\sqrt{6}}&\pi^+&K^+\\
        \pi^-&-\frac{1}{\sqrt{2}}\pi^0+\frac{\eta}{\sqrt{6}}&K^0\\
        K^-&\bar{K}^0&-\frac{2\eta}{\sqrt{6}}
\end{array}\right),\\
\mathbb{V}&=&\left(\begin{array}{ccc}
\frac{\rho^{0}}{\sqrt{2}}+\frac{\omega}{\sqrt{2}}&\rho^{+}&K^{*+}\\
\rho^{-}&-\frac{\rho^{0}}{\sqrt{2}}+\frac{\omega}{\sqrt{2}}&K^{*0}\\
K^{*-}&\bar{K}^{*0}&\phi
\end{array}\right).
\end{eqnarray}

Thus, Eqs. (\ref{eq:lag})-(\ref{eq:lag2}) can be further expanded
as follows:
\begin{eqnarray}\label{eq:lag-p-exch}
\mathcal{L}_{\tilde{\mathcal{P}}^*\tilde{\mathcal{P}}^*\mathbb{P}} &=&
i\frac{2g}{f_\pi}\varepsilon_{\alpha\mu\nu\lambda}
v^\alpha\tilde{\mathcal{P}}^{*\mu\dag}_{a}{\tilde{\mathcal{P}}}^{*\lambda}_{b}
\partial^\nu\mathbb{P}_{ab}
% +i \frac{2g}{f_\pi}\varepsilon_{\alpha\mu\beta\nu}
%v^\alpha\widetilde{\mathcal{P}}^{*\nu\dag}_{a}\widetilde{\mathcal{P}}^{*\mu}_{b}
%\partial^\beta{}\mathbb{P}_{ab}
,\label{ppp}\\
\mathcal{L}_{\tilde{\mathcal{P}}^*\tilde{\mathcal{P}}\mathbb{P}} &=&
\frac{2g}{f_\pi}(\tilde{\mathcal{P}}^{*\dag}_{a\lambda}\tilde{\mathcal{P}}_{b}+
\tilde{\mathcal{P}}^{\dag}_{a}\tilde{\mathcal{P}}^*_{b\lambda})\partial^\lambda{}\mathbb{P}_{ab}
%+\frac{2g}{f_\pi}(\widetilde{\mathcal{P}}^{*\dag}_{a\lambda}\widetilde{\mathcal{P}}^{*}_b+
%\widetilde{\mathcal{P}}^{\dag}_{a}\widetilde{\mathcal{P}}^{*}_{b\lambda})\partial^\lambda{}\mathbb{P}_{ab}
,\\
%\end{eqnarray}
%\begin{eqnarray}\label{eq:lag-v-exch}
  \mathcal{L}_{\mathcal{\tilde{P}\tilde{P}}\mathbb{V}}
  &=& \sqrt{2}\beta{}g_V\tilde{\mathcal{P}}^{\dag}_a\tilde{\mathcal{P}}_b
  v\cdot\mathbb{V}_{ab}
% +\sqrt{2}\beta{}g_V\widetilde{\mathcal{P}}^{\dag}_a
%  \widetilde{\mathcal{P}}^{}_b
%  v\cdot\mathbb{V}_{ab}
,\label{ppv}\\
  \mathcal{L}_{\tilde{\mathcal{P}}^*\tilde{\mathcal{P}}\mathbb{V}}
  &=&- 2\sqrt{2}\lambda{}g_V v^\lambda\varepsilon_{\lambda\mu\alpha\beta}
  (\tilde{\mathcal{P}}^{*\mu\dag}_a\tilde{\mathcal{P}}_b \nonumber\\&&+
  \tilde{\mathcal{P}}_a^{\dag}\tilde{\mathcal{P}}^{*\mu}_b)%\nonumber\\&&\times
  (\partial^\alpha{}\mathbb{V}^\beta)_{ab}%\\
%
%&&+  2\sqrt{2}\lambda{}g_V v^\lambda\varepsilon_{\lambda\alpha\beta\mu}
%(\widetilde{\mathcal{P}}^{*\mu\dag}_a\widetilde{\mathcal{P}}^{}_b
%+ \widetilde{\mathcal{P}}^{\dag}_a\widetilde{\mathcal{P}}_b^{*\mu})
%  (\partial^\alpha{}\mathbb{V}^\beta)_{ab}
,\nonumber\\\label{ppsv}\\
  \mathcal{L}_{\tilde{\mathcal{P}}^*\tilde{\mathcal{P}}^*\mathbb{V}}
  &=& -\sqrt{2}\beta{}g_V \tilde{\mathcal{P}}^{*\dag}_a\cdot\tilde{\mathcal{P}}_b^{*}
  v\cdot\mathbb{V}_{ab}\nonumber\\&&
  -i2\sqrt{2}\lambda{}g_V\tilde{\mathcal{P}}^{*\mu\dag}_a\tilde{\mathcal{P}}^{*\nu}_b
  (\partial_\mu{}
  \mathbb{V}_\nu - \partial_\nu{}\mathbb{V}_\mu)_{ab}%\nonumber\\
%  && - \sqrt{2}\beta g_V
%  \widetilde{\mathcal{P}}^{*\dag}_a\widetilde{\mathcal{P}}_b^{*}
%  v\cdot\mathbb{V}_{ab}
%  -i2\sqrt{2}\lambda{}g_V\widetilde{\mathcal{P}}^{*\nu\dag}_a\widetilde{\mathcal{P}}^{*\mu}_b(\partial_\mu{}
%  \mathbb{V}_\nu - \partial_\nu{}\mathbb{V}_\mu)_{ab}
,\nonumber\\\\\label{pspsv}
%  \end{eqnarray}
%\begin{eqnarray}\label{eq:lag-s-exch}
  \mathcal{L}_{\mathcal{\tilde{P}\tilde{P}}\sigma}
  &=& -2g_s\tilde{\mathcal{P}}^{}_b\tilde{\mathcal{P}}^{\dag}_b\sigma
% -2g_s\widetilde{\mathcal{P}}^{}_b\widetilde{\mathcal{P}}^{\dag}_b\sigma
,\\
  \mathcal{L}_{\tilde{\mathcal{P}}^*\tilde{\mathcal{P}}^*\sigma}
  &=& 2g_s\tilde{\mathcal{P}}^{*}_b\cdot{}\tilde{\mathcal{P}}^{*\dag}_b\sigma
 %+2g_s\widetilde{\mathcal{P}}^{*}_b\cdot{}\widetilde{\mathcal{P}}^{*\dag}_b\sigma
 .\label{sps}
\end{eqnarray}

The effective Lagrangians depicting the S-wave heavy flavor
baryons with the light mesons with chiral symmetry, heavy quark
limit and hidden local symmetry are\cite{Liu:2011xc}
\begin{eqnarray}
{\cal
L}_{\mathcal{B}_{\bar{3}}}&=&\frac12\langle\bar{\mathcal{B}}_{\bar{3}}(iv\cdot
D)\mathcal{B}_{\bar{3}}\rangle\nonumber\\&&+
i\beta_B\langle\bar{\mathcal{B}}_{\bar{3}}v^\mu(\mathcal{V}_\mu-\rho_\mu)
B_{\bar{3}}\rangle\nonumber\\&&
+\ell_B\langle\bar{\mathcal{B}}_{\bar{3}}{\sigma} \mathcal{B}_{\bar{3}}\rangle,\label{B3}\\
{\cal L}_{S}&=&-\langle\bar{\mathcal{S}}^\alpha(i v\cdot
D-\Delta_B)\mathcal{S}_\alpha\rangle\nonumber\\&&-
\frac{3}{2}g_1\epsilon^{\mu\nu\lambda\kappa}\,v_\kappa\,\langle\bar{\mathcal{S}}_\mu
A_\nu \mathcal{S}_\lambda\rangle\nonumber\\
&&+i\beta_S\langle\bar{\mathcal{S}}_\mu v_\alpha
(\mathcal{V}^\alpha-\rho^\alpha) \mathcal{S}^\mu\rangle \nonumber\\&&+
\lambda_S\langle\bar{\mathcal{S}}_\mu
F^{\mu\nu}(\rho)\mathcal{S}_\nu\rangle
\nonumber\\&&+\ell_S\langle\bar{\mathcal{S}}_\mu \sigma
\mathcal{S}^\mu\rangle .\label{B6}
\end{eqnarray}
Here, $\mathcal{S}_{\mu}^{ab}$ is composed of Dirac spinor
operators
\begin{eqnarray}
\mathcal{S}^{ab}_{\mu}&=&-\sqrt{\frac{1}{3}}(\gamma_{\mu}+v_{\mu})
    \gamma^{5}\mathcal{B}_6^{ab}+\mathcal{B}^{*ab}_{6\mu},\\
    \bar{\mathcal{S}}^{ab}_{\mu}&=&\sqrt{\frac{1}{3}}\bar{\mathcal{B}}_6^{ab}
    \gamma^{5}(\gamma_{\mu}+v_{\mu})+\bar{\mathcal{B}}^{*ab}_{6\mu}.
\end{eqnarray}
{$\mathcal{V}_\mu=\frac{1}{2}(\xi^\dag\partial_\mu\xi+\xi
\partial_\mu \xi^\dag)=\frac{i}{2f_\pi^2}[{\mathbb
P},\partial_\mu{\mathbb P}]+\cdots$}. In the above expressions,
$\mathcal{\mathcal{B}}_{\bar{3}}$ and $\mathcal{\mathcal{B}}_{6}$
denote the multiplets with $J^P=1/2^+$ in $\bar{3}_F$ and $6_F$
flavor representations respectively, while $\mathcal{B}_{6^*}$ is
the multiplet with $J^P=3/2^+$ in $6_F$ flavor representation.
Here, the $\mathcal{B}_{\bar{3}}$ and $\mathcal{B}_6$ matrices are
\begin{eqnarray}
\mathcal{B}_{\bar{3}}&=&\left(\begin{array}{ccc}
0&\Lambda^+_c&\Xi_c^+\\
-\Lambda_c^+&0&\Xi_c^0\\
-\Xi^+_c &-\Xi_c^0&0
\end{array}\right),\\
\mathcal{B}_6&=&\left(\begin{array}{ccc}
\Sigma_c^{++}&\frac{1}{\sqrt{2}}\Sigma^+_c&\frac{1}{\sqrt{2}}\Xi'^+_c\\
\frac{1}{\sqrt{2}}\Sigma^+_c&\Sigma_c^0&\frac{1}{\sqrt{2}}\Xi'^0_c\\
\frac{1}{\sqrt{2}}\Xi'^+_c&\frac{1}{\sqrt{2}}\Xi'^0_c&\Omega^0_c
\end{array}\right).
\end{eqnarray}
Additionally, $D_\mu\mathcal{B}_{\bar{3}}=\partial_\mu
\mathcal{B}_{\bar{3}} +\mathcal{V}_\mu
\mathcal{B}_{\bar{3}}+\mathcal{B}_{\bar{3}}\mathcal{V}_\mu^T$ and
$D_\mu\mathcal{S}_\nu=\partial_\mu \mathcal{S}_{\nu}
+\mathcal{V}_\mu
\mathcal{S}_{\nu}+\mathcal{S}_{\nu}\mathcal{V}_\mu^T$.

With Eqs. (\ref{B3})-(\ref{B6}), we obtain the explicit effective
Lagrangians
\begin{eqnarray}
    {\cal L}_{\mathcal{B}_{\bar{3}}\mathcal{B}_{\bar{3}}\mathbb{V}}
     &=&\frac{\beta_Bg_{{V}}}{\sqrt{2}}~\langle\bar{\mathcal{B}}_{\bar{3}}~v\cdot{\mathbb{V}} ~\mathcal{B}_{\bar{3}}\rangle,\label{ha} \\
    {\cal L}_{\mathcal{B}_{\bar{3}}\mathcal{B}_{\bar{3}}\sigma}
&=&\ell_B~\langle\bar{\mathcal{B}}_{\bar{3}} ~{\sigma}~\mathcal{B}_{\bar{3}}\rangle, \\
{\cal L}_{\mathcal{B}_6\mathcal{B}_6\mathbb{P}}
&=&\frac{ig_1}{2f_\pi}~\epsilon^{\mu\nu\lambda\kappa}v_\kappa
    \langle\bar{\mathcal{B}}_6~\gamma_\mu \gamma_\lambda
    \partial_\nu\mathbb{P} ~\mathcal{B}_6\rangle,\\
{\cal L}_{\mathcal{B}_6\mathcal{B}_6\mathbb{V}} &=& -\frac{\beta_S
g_{V}}{\sqrt{2}}~\langle\bar{\mathcal{B}}_6~ v\cdot
\mathbb{V}~ \mathcal{B}_6\rangle\nonumber\\
&&-\frac{i\lambda_Sg_{V}}{3\sqrt{2}}~\langle\bar{\mathcal{B}}_6\gamma_\mu
\gamma_\nu
(\partial^\mu \mathbb{V}^\nu-\partial^\nu\mathbb{V}^\mu)\mathcal{B}_6\rangle,\nonumber\\\\
{\cal L}_{\mathcal{B}_6\mathcal{B}_6\sigma} &=&
-\ell_S\langle\bar{\mathcal{B}}_6~\sigma~
\mathcal{B}_6\rangle.\label{ha1}
\end{eqnarray}
We list the values of the coupling constants in Eqs.
(\ref{ppp})-(\ref{sps}) and (\ref{ha})-(\ref{ha1}) in Table.
\ref{coupling}, which are given in literature\cite{Falk:1992cx,Isola:2003fh,Liu:2011xc}.

\end{multicols}

\begin{center}
\tabcaption{The parameters and coupling constants adopted in our
calculation\cite{Liu:2007bf,Falk:1992cx,Isola:2003fh,Liu:2011xc}.
\label{coupling}}
\vspace{-3mm}
\footnotesize
\begin{tabular*}{170mm}{@{\extracolsep{\fill}}ccccccccccc}
\toprule
$\beta$&$g$&$g_V$&$\lambda$ &$g_{_S}$&$\beta_B$&$\beta_S$&$\ell_B$&$\ell_S$&$g_1$&$\lambda_S$\\\hline
&&&(GeV$^{-1}$)&&&&&&&(GeV$^{-1}$)\\\midrule[1pt]
0.9&0.59&5.8&0.56 &0.76&-0.87&1.74&-3.1&6.2&0.94&3.31\\
\bottomrule
\end{tabular*}%
\end{center}

\begin{multicols}{2}

\renewcommand{\arraystretch}{1.2}
\begin{table}[hbt]
\caption{The parameters and coupling constants adopted in our
calculation \cite{Liu:2007bf,Falk:1992cx,Isola:2003fh,Liu:2011xc}.
\label{coupling}}
\begin{tabular}{cccccccccccccccccccccccccccc}\toprule[1pt]
$\beta$&$g$&$g_V$&$\lambda$ &$g_{_S}$&$\beta_B$&$\beta_S$&$\ell_B$&$\ell_S$&$g_1$&$\lambda_S$\\
&&&(GeV$^{-1}$)&&&&&&&(GeV$^{-1}$)\\\midrule[1pt]
0.9&0.59&5.8&0.56 &0.76&-0.87&1.74&-3.1&6.2&0.94&3.31\\
\bottomrule[1pt]
\end{tabular}
\end{table}

\subsection{The OBE potential}

We apply the constructed effective Lagrangians to deduce the OBE
potential of the hidden-charm molecular baryons. When calculating
the OBE potential, we first need to relate the scattering
amplitude with the OBE potential in the momentum space, which is
from the Breit approximation
\begin{eqnarray}
V(\mathbf{q})=-\frac{1}{\sqrt{\prod_i2M_i\prod_f
2M_f}}M(J,J_Z),\label{breit}
\end{eqnarray}
where $M_{i}$ and $M_f$ are the masses of the initial and final
states, respectively. Here, when deducing scattering amplitude,
the monopole form factor
$F(\mathbf{q}^2)=(\Lambda^2-m_i^2)/(\Lambda^2-q^2)$ is introduced
for compensating the off shell effect of the exchanged meson and
describing the structure effect of every interaction vertex. After
performing the Fourier transformation, we finally obtain the
effective potential in the coordinate space.

In terms of the method presented in Eq. (\ref{breit}), we obtain
the effective potentials for $\Lambda_c \bar{D}\to \Lambda_c
\bar{D}$, $\Lambda_c \bar{D}^*\to \Lambda_c \bar{D}^*$, $\Sigma_c
\bar{D}\to \Sigma_c \bar{D}$, $\Sigma_c \bar{D}^*\to \Sigma_c
\bar{D}^*$ scattering processes by exchanging $\{\omega,\sigma\}$,
$\{\omega,\sigma\}$, $\{\rho,\omega,\sigma\}$ and
$\{\pi,\eta,\rho,\omega,\sigma\}$, respectively. The corresponding
expressions of the effective potential are
\end{multicols}

\begin{center}
\begin{eqnarray}
\mathcal{V}_{\Lambda_c \bar{D}}^{I=\frac{1}{2}}(r)
&=&-2g_sl_BY(\Lambda,m_\sigma,r)-\frac{1}{2}\beta\beta_Bg_V^2Y(\Lambda,m_\omega,r),\label{hh28}\\
\mathcal{V}_{\Lambda_c \bar{D}^*}^{I=\frac{1}{2}}(r)
&=&-2g_sl_B\mbox{\boldmath$\epsilon$}_2\cdot \mbox{\boldmath$\epsilon$}_4^\dag Y(\Lambda,m_\sigma,r)-\frac{1}{2}\beta\beta_Bg_V^2\mbox{\boldmath$\epsilon$}_2\cdot \mbox{\boldmath$\epsilon$}_4^\dag Y(\Lambda,m_\omega,r),\\
\mathcal{V}_{\Sigma_c \bar{D}}^{I=\frac{3}{2}}(r)&=&-l_sg_sY(\Lambda,m_\sigma,r)+\bigg[-\frac{1}{4}\beta\beta_sg_V^2Y(\Lambda,m_\rho,r)-\frac{1}{4}\beta\beta_sg_V^2Y(\Lambda,m_\omega,r)\bigg],\\
\mathcal{V}_{\Sigma_c \bar{D}}^{I=\frac{1}{2}}(r)&=&-l_sg_sY(\Lambda,m_\sigma,r)+\bigg[\frac{1}{2}\beta\beta_sg_V^2Y(\Lambda,m_\rho,r)
-\frac{1}{4}\beta\beta_sg_V^2Y(\Lambda,m_\omega,r)\bigg],
\end{eqnarray}
\begin{eqnarray}
\mathcal{V}_{\Sigma_c \bar{D}^*}^{I=\frac{1}{2}}(r)&=&-g_sl_s\mbox{\boldmath$\epsilon$}_2\cdot \mbox{\boldmath$\epsilon$}_4^\dag Y(\Lambda,m_\sigma,r)+\bigg\{-\bigg[\frac{1}{2}\beta\beta_sg_V^2\mbox{\boldmath$\epsilon$}_2\cdot \mbox{\boldmath$\epsilon$}_4^\dag Y(\Lambda,m_\rho,r)-\frac{2\lambda \lambda_s g_V^2}{3}\bigg(-\frac{2}{3}\mbox{\boldmath $\sigma$}\cdot{\mathbf T}Z(\Lambda,m_\rho,r)\nonumber\\&&+\frac{1}{3}S(\hat{{\bf r}},\mbox{\boldmath $\sigma$},{\bf T})
T(\Lambda,m_\rho,r)\bigg)\bigg]+\frac{1}{2}\bigg[\frac{1}{2}\beta\beta_sg_V^2\mbox{\boldmath$\epsilon$}_2\cdot \mbox{\boldmath$\epsilon$}_4^\dag Y(\Lambda,m_\omega,r)-\frac{2\lambda \lambda_s g_V^2}{3}\big(-\frac{2}{3}\mbox{\boldmath $\sigma$}\cdot{\mathbf T}Z(\Lambda,m_\omega,r)\nonumber\\
&&+\frac{1}{3}S(\hat{{\bf r}},\mbox{\boldmath $\sigma$},{\bf T})T(\Lambda,m_\omega,r)\big)\bigg]\bigg\}
+\bigg\{\frac{gg_1}{f_\pi^2}\bigg[\frac{1}{3}\mbox{\boldmath $\sigma$}\cdot{\mathbf T}Z(\Lambda,m_\pi,r)+\frac{1}{3}S(\hat{{\bf r}},\mbox{\boldmath $\sigma$},{\bf T})T(\Lambda,m_\pi,r)\bigg]\nonumber\\
&&-\frac{gg_1}{6f_\pi^2}\bigg[\frac{1}{3}\mbox{\boldmath $\sigma$}\cdot{\mathbf T}Z(\Lambda,m_\eta,r)+\frac{1}{3}S(\hat{{\bf r}},\mbox{\boldmath $\sigma$},{\bf T})T(\Lambda,m_\eta,r)\bigg]\bigg\}
\label{hh31}
\end{eqnarray}
\begin{eqnarray}
\mathcal{V}_{\Sigma_c \bar{D}^*}^{I=\frac{3}{2}}(r)&=&-g_sl_s\mbox{\boldmath$\epsilon$}_2\cdot \mbox{\boldmath$\epsilon$}_4^\dag Y(\Lambda,m_\sigma,r)+\bigg\{\frac{1}{2}\bigg[\frac{1}{2}\beta\beta_sg_V^2\mbox{\boldmath$\epsilon$}_2\cdot \mbox{\boldmath$\epsilon$}_4^\dag Y(\Lambda,m_\rho,r)-\frac{2\lambda \lambda_s g_V^2}{3}\bigg(-\frac{2}{3}\mbox{\boldmath $\sigma$}\cdot{\mathbf T}Z(\Lambda,m_\rho,r)\nonumber\\&&+\frac{1}{3}S(\hat{{\bf r}},\mbox{\boldmath $\sigma$},{\bf T})
T(\Lambda,m_\rho,r)\bigg)\bigg]+\frac{1}{2}\bigg[\frac{1}{2}\beta\beta_sg_V^2\mbox{\boldmath$\epsilon$}_2\cdot \mbox{\boldmath$\epsilon$}_4^\dag Y(\Lambda,m_\omega,r)-\frac{2\lambda \lambda_s g_V^2}{3}\big(-\frac{2}{3}\mbox{\boldmath $\sigma$}\cdot{\mathbf T}Z(\Lambda,m_\omega,r)\nonumber\\
&&+\frac{1}{3}S(\hat{{\bf r}},\mbox{\boldmath $\sigma$},{\bf T})T(\Lambda,m_\omega,r)\big)\bigg]\bigg\}
+\bigg\{-\frac{gg_1}{2f_\pi^2}\bigg[\frac{1}{3}\mbox{\boldmath $\sigma$}\cdot{\mathbf T}Z(\Lambda,m_\pi,r)+\frac{1}{3}S(\hat{{\bf r}},\mbox{\boldmath $\sigma$},{\bf T})T(\Lambda,m_\pi,r)\bigg]\nonumber\\
&&-\frac{gg_1}{6f_\pi^2}\bigg[\frac{1}{3}\mbox{\boldmath $\sigma$}\cdot{\mathbf T}Z(\Lambda,m_\eta,r)+\frac{1}{3}S(\hat{{\bf r}},\mbox{\boldmath $\sigma$},{\bf T})T(\Lambda,m_\eta,r)\bigg]\bigg\}
\label{hh31}
\end{eqnarray}
\end{center}
\begin{multicols}{2}
with
\begin{eqnarray}
Y(\Lambda,m_E,r) &=& \frac{1}{4\pi r}(e^{-m_E\,r}-e^{-\Lambda r})\nonumber\\&&-\frac{\Lambda^2-m_E^2}{8\pi \Lambda }e^{-\Lambda r},\label{m1}\\
Z(\Lambda,m_E,r) &=& \bigtriangledown^2Y(\Lambda,m_E,r),\\ %= \frac{1}{r^2} \frac{\partial}{\partial r}r^2 \frac{\partial}{\partial r}Y(\Lambda,m_E,r)
T(\Lambda,m_E,r) &=&  r\frac{\partial}{\partial r}\frac{1}{r}\frac{\partial}{\partial r}Y(\Lambda,m_E,r).\label{m3}
\end{eqnarray}
Here, in the above expressions we define $S(\hat{{\bf r}},\mbox{\boldmath $\sigma$},{\bf T})=3\hat{{\mathbf r}}\cdot\mbox{\boldmath $\sigma$}
~\hat{{\mathbf r}}\cdot{\mathbf T}~-~\mbox{\boldmath $\sigma$}
\cdot{\mathbf T} $ and $\mathbf{T}=i\mbox{\boldmath$\epsilon$}_4^\dag\times \mbox{\boldmath$\epsilon$}_2$.

With effective potentials shown in Eqs. (\ref{hh28})-(\ref{hh31}),
we finally obtain the total effective potentials of the
hidden-charm systems composed of anti-charmed meson and charmed
baryon. The effective potentials shown in Eqs.
(\ref{hh28})-(\ref{hh31}) should be sandwiched between the states
in Eqs. (\ref{1})-(\ref{2}). We take the $\Sigma_c\bar{D}^*$
system with $I(\frac{3}{2}^-)$ as an example. Its total effective
potential can be expressed as
\begin{eqnarray}
V^{total}(r)=    _{1}\Big\langle
I\Big(\frac{3}{2}^-\Big)\Big|\mathcal{V}^{I=\frac{3}{2}}_{\Sigma_c\bar{D}^*}(r)\Big|I\Big(\frac{3}{2}^-\Big)\Big\rangle_1,
\end{eqnarray}
which is a three by three matrix. Using the same approach, we can
obtain the total effective potential of the other systems with
definite $I(J^P)$ quantum number. In Table. \ref{matrix}, we list
the matrixes corresponding to operators
$\mbox{\boldmath$\epsilon$}_2\cdot\mbox{\boldmath$\epsilon$}_4^\dag$,
$\mbox{\boldmath $\sigma$}\cdot{\mathbf T}$ and $S(\hat{{\bf r}},\mbox{\boldmath $\sigma$},{\bf T})$ in
Eqs.~ (\ref{hh28})-(\ref{hh31}) when transferring the potentials
in Eq. (\ref{hh28})-(\ref{hh31}) into the total effective
potentials of the hidden-charm systems composed of the
anti-charmed meson and charmed baryon.

\end{multicols}

\begin{center}
\tabcaption{The matrixes corresponding to $_1\langle
I(\frac{1}{2}^-)|\mathcal{O}_i|I(\frac{1}{2}^-)\rangle_1$ and
$_1\langle
I(\frac{3}{2}^-)|\mathcal{O}_i|I(\frac{3}{2}^-)\rangle_1$, where
$\mathcal{O}_i$ denotes operators
$\mbox{\boldmath$\epsilon$}_2\cdot\mbox{\boldmath$\epsilon$}_4^\dag$,
$\mbox{\boldmath $\sigma$}\cdot{\mathbf T}$ and $S(\hat{{\bf r}},\mbox{\boldmath $\sigma$},{\bf T})$ in
Eqs.~ (\ref{hh28})-(\ref{hh31}). Here,
$|I(\frac{1}{2}^-)\rangle_1$ and $|I(\frac{3}{2}^-)\rangle_1$ are
defined in
Eqs.~(\ref{Eq:wf1})-(\ref{Eq:wf3}).\label{matrix}}
\vspace{-3mm}
\footnotesize
\begin{tabular*}{170mm}{@{\extracolsep{\fill}}cccc}
\toprule
&
$\mbox{\boldmath$\epsilon$}_2\cdot\mbox{\boldmath$\epsilon$}_4^\dag$
&$\mbox{\boldmath $\sigma$}\cdot{\mathbf T}$& $S(\hat{{\bf r}},\mbox{\boldmath $\sigma$},{\bf T})$\\%\hline
\midrule[1pt] $_1\langle
I(\frac{1}{2}^-)|\mathcal{O}_i|I(\frac{1}{2}^-)\rangle_1$&
$\left(\begin{array}{cc}
1&0\\
0&1\\
\end{array}\right)$&
$\left(\begin{array}{cc}
2&0\\
0&-1\\
\end{array}\right)$&
$\left(\begin{array}{cc}
0&-\sqrt{2}\\
-\sqrt{2}&2\\
\end{array}\right)$\\
\midrule[1pt] $_1\langle
I(\frac{3}{2}^-)|\mathcal{O}_i|I(\frac{3}{2}^-)\rangle_1$&
$\left(\begin{array}{ccc}
1&0&0\\
0&1&0\\
0&0&1\\
\end{array}\right)$&
$\left(\begin{array}{ccc}
-1&0&0\\
0&2&0\\
0&0&-1\\
\end{array}\right)$&
$\left(\begin{array}{ccc}
0&1&-2\\
1&0&-1\\
-2&-1&0\\
\end{array}\right)$\\
\bottomrule
\end{tabular*}%
\end{center}

\begin{multicols}{2}

The kinetic terms are
\begin{eqnarray}
K_{|I(\frac{1}{2}^-)\rangle_0}&=&-\frac{\triangle}{2\tilde{m}},\\
    K_{|I(\frac{1}{2}^-)\rangle_1}&=&\mathrm{diag}\Bigg(-\frac{\triangle}{2\tilde{m}},~
    -\frac{\triangle_2}{2\tilde{m}}\Bigg),\\
    K_{|I(\frac{3}{2}^-)\rangle_1}&=&\mathrm{diag}\Bigg(-\frac{\triangle}{2\tilde{m}},~
    -\frac{\triangle_2}{2\tilde{m}},~
    -\frac{\triangle_2}{2\tilde{m}}\Bigg)
\end{eqnarray}
corresponding to the systems in Eqs.
(\ref{Eq:wf01})-(\ref{Eq:wf3}) respectively, where
$\triangle=\frac{1}{r^2}\frac{\partial}{\partial
r}r^2~\frac{\partial}{\partial r}$, $\triangle_2=\triangle
-{6\over{r^2}}$. $\tilde{m}=m_Bm_{P^{(*)}}/(m_B+m_{P^{(*)}})$ is
the reduced mass of the system, where $m_B$ and $m_{P^{(*)}}$ are
the masses of charmed baryon and pseudoscalar (vector)
anti-charmed meson, respectively.

\section{Numerical results}\label{sec3}

In this work, we mainly investigate the hidden-charm systems
$\Lambda_c \bar{D}$ with $\frac{1}{2}(\frac{1}{2}^-)$, $\Lambda_c
\bar{D}^*$ with
$\frac{1}{2}(\frac{1}{2}^-),\frac{1}{2}(\frac{3}{2}^-)$, $\Sigma_c
\bar{D}$ with
$\frac{1}{2}(\frac{1}{2}^-),\frac{3}{2}(\frac{1}{2}^-)$, $\Sigma_c
\bar{D}^*$ with
$\frac{1}{2}(\frac{1}{2}^-),\frac{3}{2}(\frac{3}{2}^-),\frac{1}{2}(\frac{1}{2}^-),\frac{3}{2}(\frac{3}{2}^-)$.
If we replace $\bar{D}^{(*)}$ and charmed baryon by the
corresponding $B^{(*)}$ and bottom baryon, we can extend the same
formalism listed to discuss the hidden-bottom molecular baryons
composed of a bottom meson and a bottom baryon, which include
$\Lambda_b B$ with $\frac{1}{2}(\frac{1}{2}^-)$, $\Lambda_b B^*$
with $\frac{1}{2}(\frac{1}{2}^-),\frac{1}{2}(\frac{3}{2}^-)$,
$\Sigma_b B$ with
$\frac{1}{2}(\frac{1}{2}^-),\frac{3}{2}(\frac{1}{2}^-)$, $\Sigma_b
B^*$ with $\frac{1}{2}(\frac{1}{2}^-),\frac{3}{2}(\frac{3}{2}^-),
\frac{1}{2}(\frac{1}{2}^-),\frac{3}{2}(\frac{3}{2}^-)$.

Using the potential obtained above, the binding energy can be
obtained by solving the coupled-channel Schr\"odinger equation. We
use the FESSDE program \cite{Abrashkevich1995} to produce the
numerical results for the binding energy and the corresponding
root-mean-square radius $r$ with the variation of the cutoff
$\Lambda$ in the region of $0.8\leq\Lambda\leq2.2$ GeV as shown in
Fig.~\ref{Fig:E}. Here, we only show the bound state solution with
binding energy less than 50 MeV since the OBE model is only valid
to deal with the loosely bound hadronic molecular system.

One notices that $\Lambda_c$ does not combine with $\bar{D}^{(*)}$
to form a hidden-charm molecular state. There does exist a
hidden-bottom molecular state composed of $\Lambda_b$ and
$B^{(*)}$. As shown in Fig.~\ref{Fig:E}, we find bound state
solutions only for five hidden-charm states, i.e., $\Sigma_c\bar{D}^*$ states with $I(J^P)=\frac{1}{2}(\frac{1}{2}^-), \frac{1}{2}(\frac{3}{2}^-), \frac{3}{2}(\frac{1}{2}^-), \frac{3}{2}(\frac{3}{2}^-)$ and $\Sigma_c\bar{D}$ state with $\frac{3}{2}(\frac{1}{2}^-)$.
We also find the bound state solutions for the hidden-bottom
molecular baryons, which are $\Sigma_bB^*$ states with
$I(J^P)=\frac{1}{2}(\frac{1}{2}^-), \frac{1}{2}(\frac{3}{2}^-), \frac{3}{2}(\frac{1}{2}^-), \frac{3}{2}(\frac{3}{2}^-)$ and $\Sigma_b B$ with
$\frac{3}{2}(\frac{1}{2}^-)$.

\end{multicols}

\begin{center}
\includegraphics[bb=160 140 550 400,clip,scale=0.6]{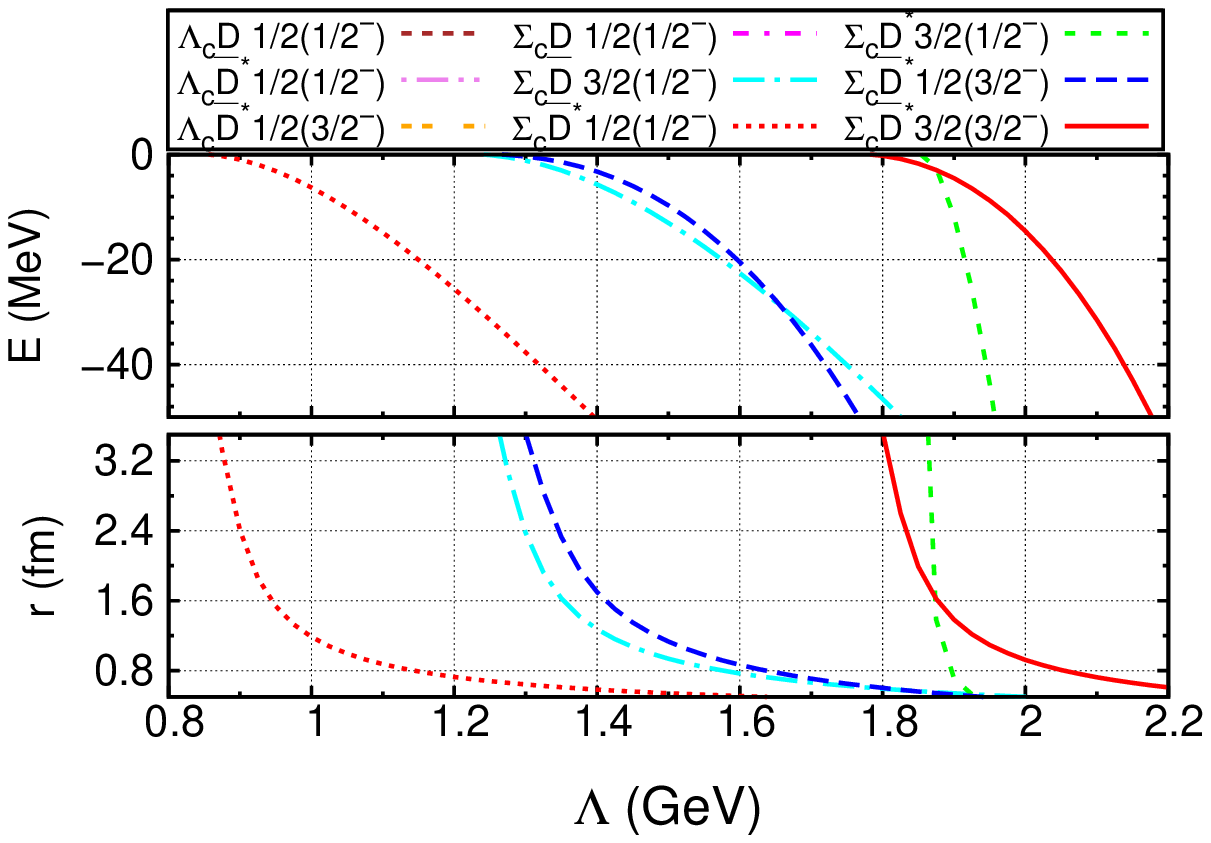}
\includegraphics[bb=160 140 550 400,clip,scale=0.6]{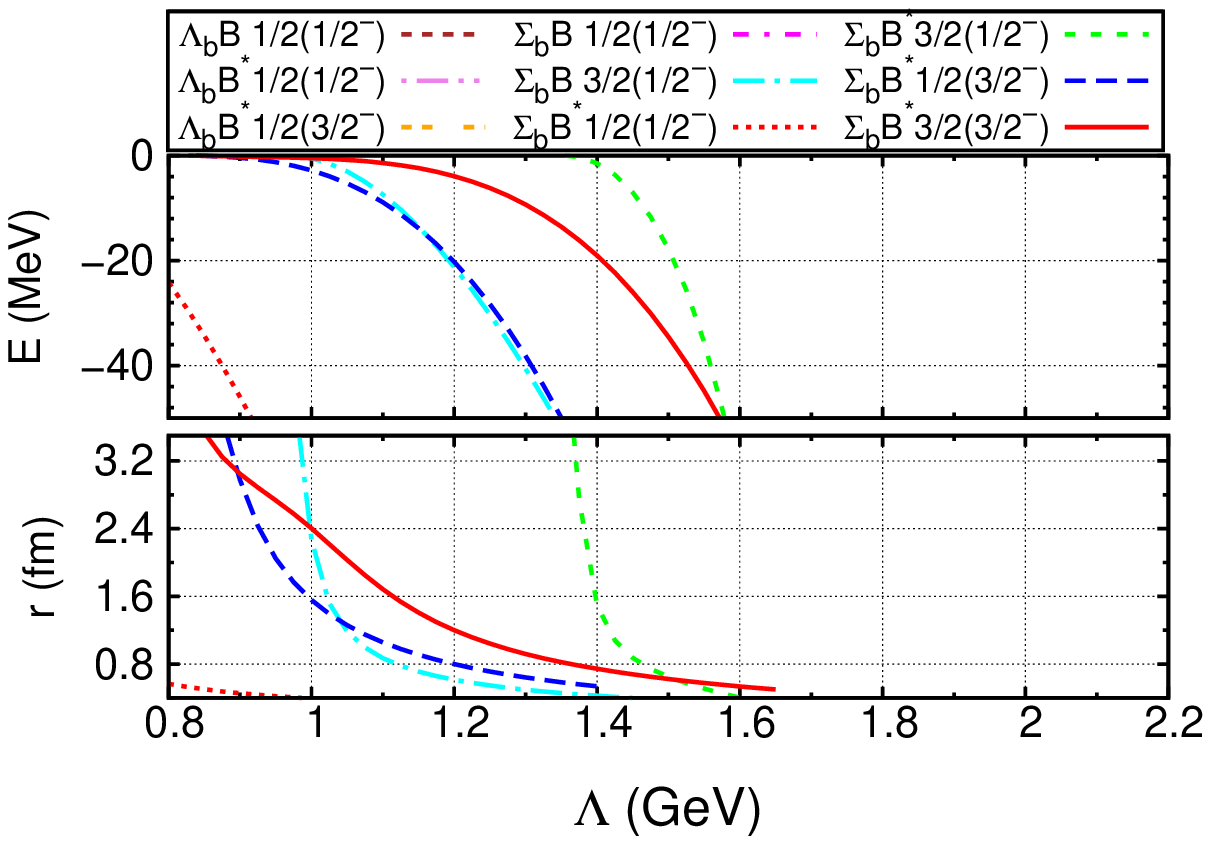}
\figcaption{(Color online). The $\Lambda$ dependence of the binding
energy and the obtained root-mean-square radius $r$ of the
hidden-charm or hidden-bottom system. \label{Fig:E}}
\end{center}

\begin{center}
\includegraphics[bb=-50 410 800 680,clip,scale=0.8]{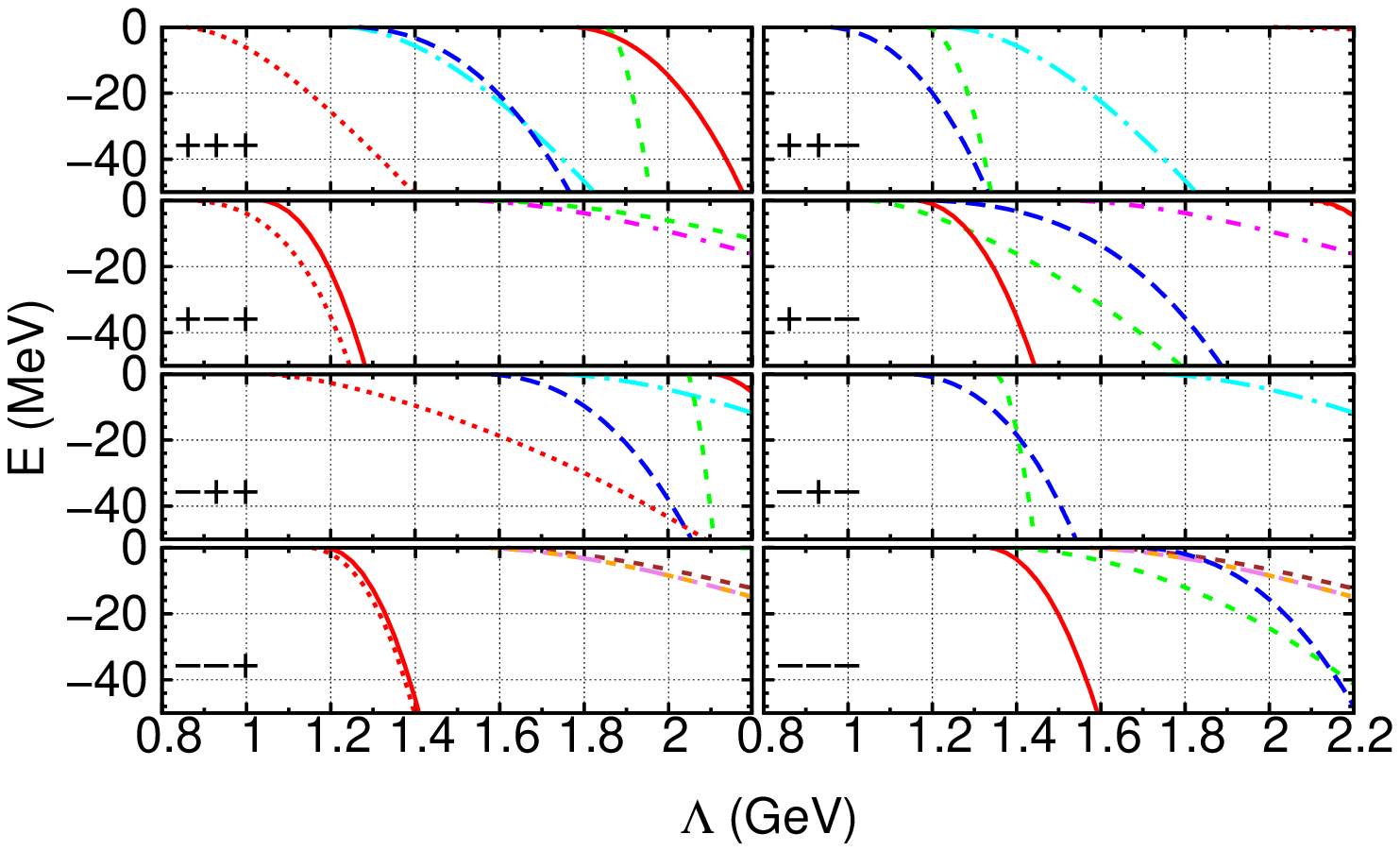}\\
\includegraphics[bb=-50 410 800 680,clip,scale=0.8]{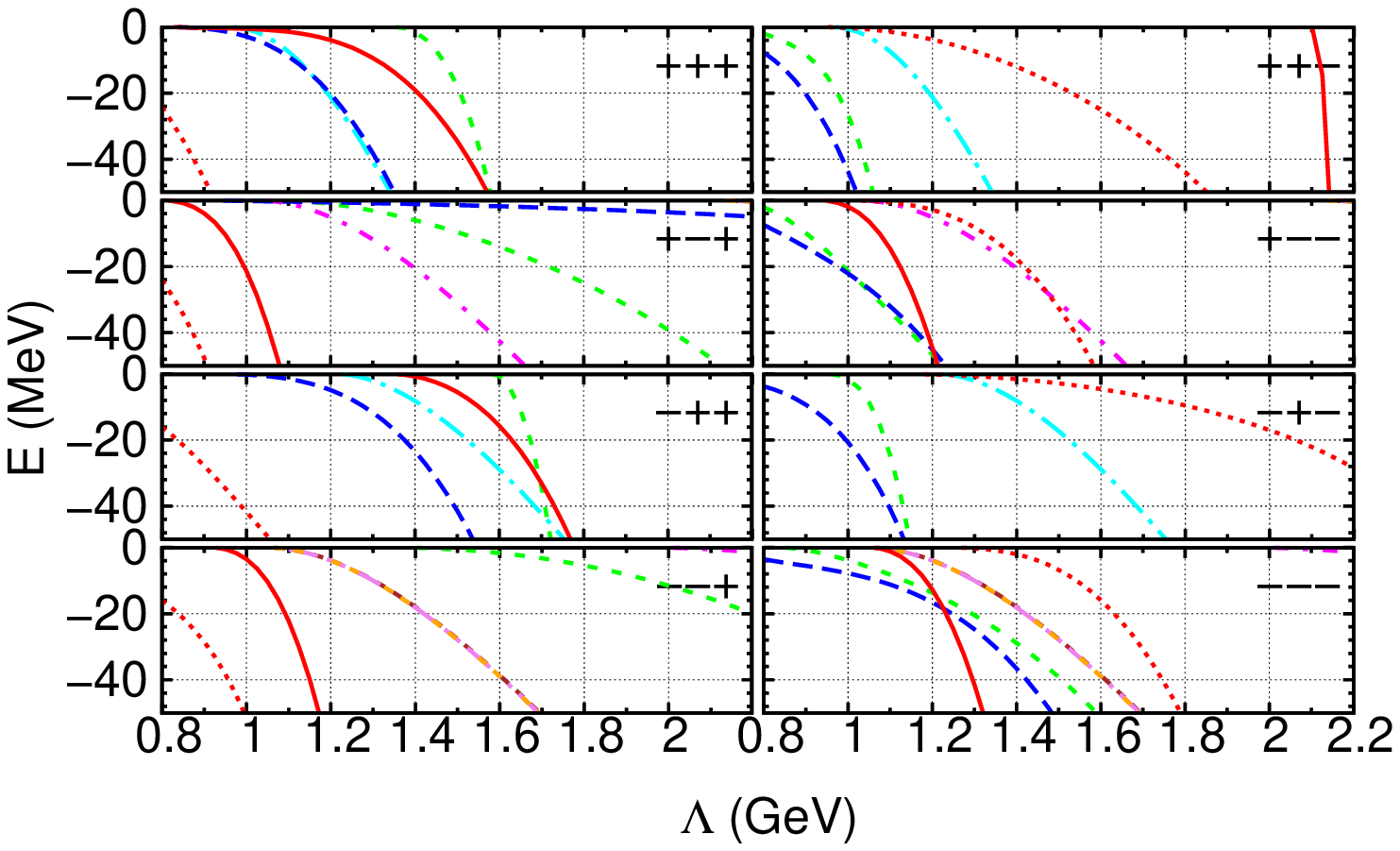}\\
\figcaption{(Color online). The binding energy of the hidden-charm state (top) or
hidden-bottom state (bottom). Here, $+/-$ in $``\pm1\pm1\pm1"$ denotes
that we need to multiply the corresponding, sigma, vector and pion
exchange potentials by an extra factor $+1/-1$, which come from
the changes of the signs of the coupling constants. The other
conventions are the same as in Fig. \ref{Fig:E}. \label{Fig:sign}}
\end{center}

\begin{multicols}{2}

For the heavy baryon sector, we adopt the values of coupling
constants including the signs  as given in Ref. \cite{Liu:2011xc}.
However, for the heavy meson sector, the signs of the coupling
constants $g$, $\beta/\lambda$, $g_s$, can not be well constrained
by the available experimental data or theoretical considerations,
which results in uncertainty of the signs of the corresponding
sigma, vector and pion exchange potentials. For the sake of
completeness, we present the dependence of the binding energy on
$\Lambda$ under eight combinations of the signs of $g$,
$\beta/\lambda$, $g_s$ as shown in Fig.~\ref{Fig:sign}. The
notation $+/-$ denotes an extra factor $+1/-1$ which changes the
signs of $g$, $\beta/\lambda$, $g_s$ in the corresponding pion,
vector and sigma exchange potentials. Generally speaking, the
sigma exchange contribution is negligible while the $\pi$ and
$\rho/\omega$ meson exchanges play a very important role.

\section{Discussion and conclusion}\label{sec4}

In this work, we have employed the OBE model to study whether
there exist the loosely bound hidden-charm molecular states
composed of an S-wave anti-charmed meson and an S-wave charmed
baryon. Our numerical results indicate that there do not exist
$\Lambda_c \bar{D}$ and $\Lambda_c \bar{D}^*$ molecular states due
to the absence of bound state solution, which is an interesting
observation in this work. Additionally, we notice the bound state
solutions only for five hidden-charm states, i.e., $\Sigma_c\bar{D}^*$ states with $I(J^P)=\frac{1}{2}(\frac{1}{2}^-), \frac{1}{2}(\frac{3}{2}^-), \frac{3}{2}(\frac{1}{2}^-), \frac{3}{2}(\frac{3}{2}^-)$ and $\Sigma_c\bar{D}$ state with $\frac{3}{2}(\frac{1}{2}^-)$. We also extend the same formulism to
study hidden-bottom system with an S-wave bottom meson and an
S-wave bottom baryon. The mass of the component in the
hidden-bottom system is heavier than that in the hidden-charm
system, which leads to the reduced kinetic energy and is helpful
to the formation of the loosely bound states. Our numerical
results have confirmed this point. There exist the $\Sigma_bB^*$ molecular states with
$I(J^P)=\frac{1}{2}(\frac{1}{2}^-), \frac{1}{2}(\frac{3}{2}^-), \frac{3}{2}(\frac{1}{2}^-), \frac{3}{2}(\frac{3}{2}^-)$ and $\Sigma_b B$ state with
$\frac{3}{2}(\frac{1}{2}^-)$.

The hidden-charm systems composed of an S-wave anti-charmed meson
and an S-wave charmed baryon are very interesting. Since the
masses of such exotic systems are around 4 GeV, they may be
accessible to the forthcoming PANDA, Belle-II and SuperB
experiments. These exotic hidden-bottom baryons might be searched
for at J-PARC or LHCb. The exploration of these states may shed
light on the mechanism of forming molecular states and help reveal
underlying structures of some of those newly observed
near-threshold hadrons.

\end{multicols}

\vspace{-1mm}
\centerline{\rule{80mm}{0.1pt}}
\vspace{2mm}

\begin{multicols}{2}

\end{multicols}

\clearpage

%\end{CJK*}
\end{document}